\documentclass[
  reprint,
  amsmath,amssymb,
  aps,
  prb,
  superscriptaddress,
  longbibliography
]{revtex4-2}

\usepackage{graphicx}
\usepackage{bm}
\usepackage{xcolor}
\usepackage{hyperref}

\hypersetup{
  colorlinks=true,
  linkcolor=blue,
  citecolor=blue,
  urlcolor=blue
}

\newcommand{\dd}{\mathrm{d}}
\newcommand{\ii}{\mathrm{i}}
\newcommand{\rvec}{\mathbf r}
\newcommand{\qvec}{\mathbf q}
\newcommand{\Gvec}{\mathbf G}
\newcommand{\Xvec}{\mathbf X}
\newcommand{\uvec}{\mathbf u}
\newcommand{\Acell}{A_{\rm cell}}
\newcommand{\Ang}{\text{\AA}}
\newcommand{\taueq}{\tau_{\rm eq}}
\newcommand{\taudep}{\tau_{\rm dep}}

\begin{document}

\title{A Thermodynamic-Limit Pinning Criterion for Two-Dimensional Structural Superlubricity}

\author{Li Wang}
\email{wangli233@smail.nju.edu.cn}
\affiliation{School of Physics, Nanjing University, Nanjing 210093, China}

\author{Yunjie Ye}
\affiliation{Sichuan ZeroNestor Microelectronics Technology Co., Ltd., Chengdu, Sichuan, China}

\date{July 22,2026}

\begin{abstract}
Incommensurability and elastic reconstruction do not by themselves define a structurally superlubric phase. We define fully sliding and pinned zero-temperature phases by $\limsup_{A\to\infty}\tau_{\rm dep}^{\max}(A)=0$ and $\liminf_{A\to\infty}\tau_{\rm dep}^{\min}(A)>0$, respectively; $\Lambda_n=|V_n|G_{n,i}[D_{\rm rel}^{-1}(\mathbf q_n)]_{ij}G_{n,j}$ measures only reconstruction susceptibility. Translational covariance then proves that a clean, smooth, infinite moir\'e continuum can reconstruct without acquiring a bulk sliding barrier. We restore atomic sampling in a two-dimensional discrete model of graphene/hBN and test both a diffusion quantum Monte Carlo first-star potential and a 15-harmonic Leven potential across three rational approximants and five directions. No physical-coupling equilibrium or metastable barrier is resolved. The Leven spectrum raises the largest tested $\Lambda$ from $0.142$ to $0.212$, while artificial scaling through $\Lambda=1$ reaches uncontrolled strain before a size-independent threshold appears. The tested zero-temperature in-plane models are therefore consistent with an elastically relaxed sliding regime; $\Lambda=1$ is a reconstruction scale, not a static phase criterion.
\end{abstract}

\maketitle

\section{Introduction}

Structural superlubricity is the suppression of static friction by cancellation of lateral forces across a crystalline interface. It was anticipated from atomistic locking arguments and from the Aubry transition of incommensurate Frenkel--Kontorova systems \cite{Hirano1990,Aubry1983,MuserRobbins2000,Muser2004}. Experiments on graphite, graphene nanoribbons, and layered heterojunctions have since found ultralow friction from nanometer contacts to a single submillimeter graphite contact \cite{Dienwiebel2004,Zheng2008,Liu2012,Yang2013,Dietzel2013,Kawai2016,Song2018,Liao2022,Wang2020Characterization,Han2026Macroscale}. These observations also make clear that load, elastic compliance, edges, contamination, and grain boundaries can determine whether low friction survives in a real contact \cite{Sharp2016,FengXu2022,Vanossi2013,Hod2018,Vanossi2020,Wang2024RMP}.

Edge control has recently brought this question directly into the microscale regime: selective removal of amorphous edges was reported to eliminate the static-friction peak in graphite contacts with areas up to $100~\mu{\rm m}^2$ \cite{Li2024ZeroStatic}. In parallel, the ``slidevice'' framework organizes generators, resonators, sliding ferroelectrics, and reconfigurable components whose operation relies on structurally superlubric interfaces \cite{Peng2025Slidevices}. These developments sharpen the need to distinguish a bulk phase criterion from the performance of a finite device.

Geometric incommensurability is not, by itself, a phase criterion. An incommensurate interface may reconstruct strongly, and a finite or polycrystalline contact may retain edge or defect friction even when forces cancel in its interior \cite{Varini2015,Gigli2017,Minkin2021,Qu2020,Hu2024}. The converse is equally important: commensurate domains do not imply a nonzero thermodynamic shear strength. Graphene on hexagonal boron nitride (hBN) exposes both points. Near alignment it forms reconstructed domains separated by solitons \cite{Woods2014,SanJose2014,Carr2018}, while aligned graphite/hBN contacts can nevertheless exhibit structural superlubricity \cite{Song2018}.

Existing graphene/hBN theories resolve different parts of this problem. The registry-index model of Leven \emph{et al.} predicts strong cancellation of the rigid-flake corrugation over most orientations \cite{Leven2013}. Leven \emph{et al.} subsequently developed a registry- and separation-dependent interlayer potential \cite{Leven2016}. Mandelli \emph{et al.} used that description in finite-flake simulations with elastic and out-of-plane deformation, normal load, and dissipation, finding a moir\'e-controlled crossover to low friction even near alignment \cite{Mandelli2017}. Yan \emph{et al.} showed that finite rigid sliders have shape- and edge-dependent static-force oscillations rather than a simple area law \cite{YanGao2024Shape}. Gao \emph{et al.} subsequently traced kinetic dissipation to elastic moir\'e pinning at corners and edges and identified corner-, edge-, and surface-dominated regimes \cite{Gao2025Edge}. Those descriptions are appropriate for finite contacts. What they do not supply is a thermodynamic criterion that distinguishes subextensive boundary forces from a nonzero bulk depinning stress.

The distinction also separates two transitions that are often treated as interchangeable. The Pokrovsky--Talapov commensurate--incommensurate transition marks the energetic onset of solitons or domains in a smooth elastic medium \cite{PokrovskyTalapov1979,Lazarides2009}. The Aubry transition marks the onset of a finite depinning threshold in an incommensurate discrete system \cite{Aubry1983,Mandelli2015,Brazda2018}. Both compare adhesion with elasticity, but their critical parameters need not coincide. Driven stick--slip is a third problem because its onset also depends on the loading stiffness \cite{Socoliuc2004,Wang2024StickSlip}.

We separate reconstruction from pinning in two steps. The static phase is defined by the thermodynamic scaling of the directional depinning stress; the relaxed equilibrium corrugation is retained as an independent diagnostic. Reconstruction is quantified by the tensor susceptibility
\begin{equation}
\Lambda_n
=
|V_n|\,G_{n,i}
\left[
D_{\rm rel}^{-1}(\qvec_n)
\right]_{ij}
G_{n,j}
\label{eq:lambda_tensor_intro}
\end{equation}
where \(V_n\) is a generalized stacking-fault-energy coefficient per area, \(\Gvec_n\) is its stacking reciprocal vector, \(\qvec_n\) is the associated moir\'e wave vector, and \(D_{\rm rel}\) is the relative elastic kernel. The quantity \(\Lambda_n\) measures elastic response to a registry harmonic. It is not an order parameter and has no universal critical value for pinning.

We apply this separation to graphene/hBN using two independently parameterized adhesion landscapes: the first reciprocal star obtained by diffusion quantum Monte Carlo (DMC) \cite{Szyniszewski2025}, and 15 reciprocal harmonics extracted from the registry-dependent Leven interlayer potential \cite{Leven2016}. The latter increases the largest tested susceptibility from $0.142$ to $0.212$ for two free layers. We then ask the phase-defining question in a two-dimensional discrete Frenkel--Kontorova model: does a barrier survive increasing rational approximants and different loading directions? For neither landscape do we resolve such a barrier at physical coupling. Two-dimensional shift-torus and multi-start calculations independently test equilibrium corrugation and metastability. These finite calculations support a sliding-side assignment for the specified zero-temperature, defect-free, in-plane models; they do not prove the infinite-size limit or address kinetic friction, thermal activation, wear, plasticity, contamination, or electronic and phononic dissipation.

\section{Relaxed corrugation and phase definition}

Consider two crystalline layers forming an interface over area \(A\).
The imposed global lateral displacement is \(\Xvec\), and the slowly varying internal relative displacement field is \(\uvec(\rvec)\).
Let \(B\) be the mismatch matrix that maps real-space position to local stacking displacement. For small isotropic mismatch \(\delta\) and twist angle \(\theta\), we use \(B\simeq\delta I+\theta\epsilon\), where \(I\) is the two-dimensional identity and \(\epsilon\mathbf v=\hat{\mathbf z}\times\mathbf v\) for any in-plane vector \(\mathbf v\).
At fixed \(\Xvec\), the energy functional is
\begin{equation}
E[\uvec,\Xvec;A]
=
E_{\rm el}[\uvec;A]
+
\int_A \dd^2 r\,
V_{\rm GSFE}
\!\left(
B\rvec+\Xvec+\uvec(\rvec)
\right),
\label{eq:energy_functional}
\end{equation}
where \(V_{\rm GSFE}\) is the generalized stacking-fault energy density, periodic in its displacement argument.
The decomposition into \(\Xvec\) and \(\uvec\) contains a translational zero mode.  We fix it by imposing
\begin{equation}
\langle\uvec\rangle_A
=
\frac{1}{A}\int_A\dd^2r\,\uvec(\rvec)
=0.
\label{eq:zero_mode_constraint}
\end{equation}
Without this constraint, a uniform component of \(\uvec\) could absorb \(\Xvec\), and the sliding landscape would be undefined.  The constrained relaxed sliding energy is
\begin{equation}
E_{\min}(\Xvec;A)
=
\min_{\uvec(\rvec):\langle\uvec\rangle_A=0}
E[\uvec,\Xvec;A].
\label{eq:emin}
\end{equation}
Its equilibrium restoring force and stress are
\begin{equation}
F_{\rm eq}(A)
=
\max_{\Xvec}
\left|
\nabla_{\Xvec}E_{\min}(\Xvec;A)
\right|,
\label{eq:Feq}
\end{equation}
and
\begin{equation}
\taueq(A)
=
\frac{F_{\rm eq}(A)}{A}.
\label{eq:tau_eq}
\end{equation}
For a unit in-plane direction $\hat{\mathbf e}$, we also define the directional equilibrium stress
\begin{equation}
\tau_{\rm eq}(A,\hat{\mathbf e})
=\frac{1}{A}\max_{\Xvec}
\left|\hat{\mathbf e}\cdot\nabla_{\Xvec}E_{\min}(\Xvec;A)\right|,
\label{eq:directional_tau_eq}
\end{equation}
so that $\taueq(A)=\max_{\hat{\mathbf e}}\tau_{\rm eq}(A,\hat{\mathbf e})$.

Static friction is a stability property when several local minima coexist. Let $\uvec_m(\Xvec)$ denote a continuously followed local minimum and
\begin{equation}
E_m(\Xvec;A)=E[\uvec_m(\Xvec),\Xvec;A]
\label{eq:branch_energy}
\end{equation}
its branch energy; unlike $E_{\min}$, $E_m$ need not be the global envelope. Under a shear stress \(\boldsymbol\tau\), the full tilted enthalpy is
\begin{equation}
\mathcal H_{\boldsymbol\tau}
=E-A\boldsymbol\tau\cdot\Xvec,
\label{eq:tilted_enthalpy}
\end{equation}
where \(\Xvec\) is lifted from the registry torus to the unwrapped sliding coordinate along the quasistatic branch.
Starting from a specified zero-stress minimum, \(\taudep(A,\hat{\mathbf e})\) is the smallest positive scalar stress applied along the unit vector \(\hat{\mathbf e}\) at which the continuously connected mechanically stable branch loses stability. The branch-selection rule---for example, continuation from the equilibrium minimum or from a stated preparation protocol---is part of the finite-size ensemble. Define
\begin{equation}
\tau_{\rm dep}^{\min}(A)=\min_{\hat{\mathbf e}}\taudep(A,\hat{\mathbf e}),
\qquad
\tau_{\rm dep}^{\max}(A)=\max_{\hat{\mathbf e}}\taudep(A,\hat{\mathbf e}).
\label{eq:tau_dep}
\end{equation}
In a reversible single-branch problem $E_m=E_{\min}$ and the directional depinning stress follows from the constrained corrugation. With hysteresis or branch switching, $E_m$ and $E_{\min}$ differ, so equilibrium and depinning stresses must both be reported.

We define a fully sliding (structurally superlubric) phase by
\begin{equation}
\limsup_{A\to\infty}\tau_{\rm dep}^{\max}(A)=0,
\label{eq:superlubric_definition}
\end{equation}
and a fully pinned phase by
\begin{equation}
\liminf_{A\to\infty}\tau_{\rm dep}^{\min}(A)>0.
\label{eq:pinned_definition}
\end{equation}
If there are fixed directions $\hat{\mathbf e}_{s}$ and $\hat{\mathbf e}_{p}$ such that
\begin{equation}
\limsup_{A\to\infty}\taudep(A,\hat{\mathbf e}_{s})=0,
\qquad
\liminf_{A\to\infty}\taudep(A,\hat{\mathbf e}_{p})>0,
\label{eq:direction_selective_definition}
\end{equation}
the interface is direction-selective: at least one sliding direction survives while another remains pinned. This class is absent from an isotropic scalar treatment but is natural for low-symmetry crystals. Sequences not covered by Eqs.~\eqref{eq:superlubric_definition}--\eqref{eq:direction_selective_definition}, or in which the disorder ensemble changes with area, remain unassigned until their finite-size ensemble is specified.
The limit is taken along a specified sequence of contacts, boundary conditions, and branch-selection rules, since edge-dominated contacts, coherent single crystals, polycrystalline contacts, and differently prepared metastable states need not share the same finite-size scaling.

\begin{figure*}[t]
\centering
\includegraphics[width=0.98\textwidth]{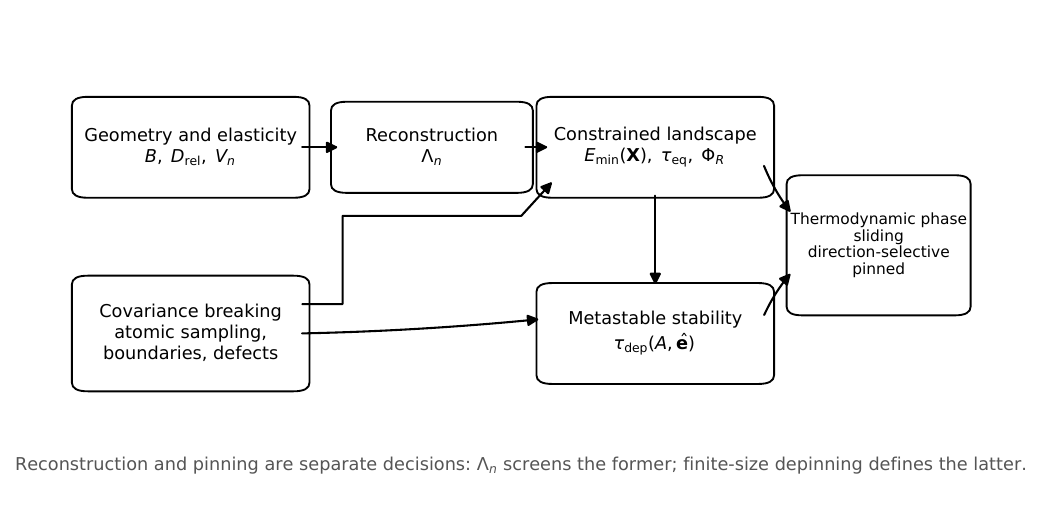}
\caption{
Logical structure of the criterion.
Bare geometry and elasticity determine the reconstruction susceptibility \(\Lambda\), whereas the thermodynamic phase is determined by the size scaling of the depinning stress \(\taudep\); \(\Phi_R\) and \(\taueq\) diagnose its equilibrium corrugation.
Reconstruction is not sufficient for bulk static pinning. Exact commensuration can produce a bulk barrier, while edges and defects can pin finite contacts.
}
\label{fig:concept}
\end{figure*}

For a finite contact, a commensurate approximant, or a moir\'e supercell, let $\Omega_X$ be one primitive cell in the two-dimensional collective-displacement space and $\mathcal L_X^*$ its reciprocal lattice. Define the sliding-cell average
\begin{equation}
\overline E(A)=\frac{1}{|\Omega_X|}
\int_{\Omega_X}\dd^2X\,E_{\min}(\mathbf X;A).
\label{eq:mean_sliding_energy}
\end{equation}
The Fourier coefficient of the relaxed sliding energy density is
\begin{equation}
\begin{split}
V^{\rm ren}_{\mathbf K}(A)
={}&\frac{1}{|\Omega_X|}
\int_{\Omega_X}\dd^2X\,
\\[-2pt]
&\times
\frac{E_{\min}(\mathbf X;A)-\overline E(A)}{A}
e^{-\ii\mathbf K\cdot\mathbf X}.
\end{split}
\label{eq:vren_projection}
\end{equation}
Here $\mathbf K\in\mathcal L_X^*$ and $|\Omega_X|$ is the area of the sliding cell. The vectors $\mathbf K$ need not equal the atomic stacking vectors $\mathbf G_n$; for a rational approximant the relaxed landscape can have a shorter primitive period.
The coherent corrugation over all retained harmonics is summarized by its root-mean-square restoring stress,
\begin{equation}
\begin{split}
\Phi_R(A)
&=
\left[
\sum_{\mathbf K\in\mathcal L_X^*\setminus\{0\}}|\mathbf K|^2
|V^{\rm ren}_{\mathbf K}(A)|^2
\right]^{1/2}
\\
&=
\left[
\frac{1}{|\Omega_X|}\int_{\Omega_X}\dd^2X\,
\left|\nabla_{\mathbf X}\frac{E_{\min}}{A}\right|^2
\right]^{1/2}.
\end{split}
\label{eq:phiR_revised}
\end{equation}
Parseval's identity gives \(\Phi_R(A)\le\taueq(A)\). Thus a nonzero limiting \(\Phi_R\) proves a nonvanishing equilibrium restoring stress in at least one direction, but not full directional pinning; conversely, \(\Phi_R\to0\) alone does not exclude narrow force spikes or metastable pinning. The phase criteria are Eqs.~\eqref{eq:superlubric_definition}--\eqref{eq:direction_selective_definition}; \(\Phi_R\) is a spectral diagnostic, not an assumed equivalent order parameter.
Figure~\ref{fig:concept} summarizes the resulting separation between reconstruction, equilibrium corrugation, and depinning.

\section{Registry coherence and continuum covariance}

Expand the generalized stacking-fault energy as
\begin{equation}
V_{\rm GSFE}(\mathbf b)
=
\sum_n
V_n e^{\ii\Gvec_n\cdot\mathbf b}
+
{\rm c.c.},
\label{eq:Vint}
\end{equation}
where \(V_n\) has units of energy per area and c.c. denotes the complex-conjugate term.
The mismatch matrix generates
\begin{equation}
\qvec_n=B^{T}\Gvec_n.
\label{eq:q_from_B}
\end{equation}
The \(n\)-th local registry harmonic is therefore
\begin{equation}
V_n
e^{\ii[
\qvec_n\cdot\rvec
+
\Gvec_n\cdot\Xvec
+
\Gvec_n\cdot\uvec(\rvec)
]}
+
{\rm c.c.}
\label{eq:moire_term}
\end{equation}
For a minimizing field \(\uvec_{\Xvec}^{\ast}\), define
\begin{equation}
\Psi_n(\Xvec,A)
=
\frac{1}{A}
\int_A \dd^2 r\,
e^{\ii[
\qvec_n\cdot\rvec
+
\Gvec_n\cdot\uvec_{\Xvec}^{\ast}(\rvec)
]}.
\label{eq:Psi}
\end{equation}
The corresponding contribution to the relaxed energy density is
\begin{equation}
V_n\Psi_n(\Xvec,A)e^{\ii\Gvec_n\cdot\Xvec}
+
{\rm c.c.}
\label{eq:coherent_contribution}
\end{equation}
The explicit \(\Xvec\) dependence of \(\Psi_n\) is essential.
Evaluating \(\Psi_n\) at one displacement and identifying \(V_n\Psi_n\) with a sliding barrier would neglect reoptimization of the displacement field.

This observation gives a useful no-go result.
For an infinite clean interface governed only by the smooth continuum functional in Eq.~\eqref{eq:energy_functional}, assume that the mismatch matrix \(B\) is nonsingular, take a shift \(\Delta\Xvec\), and choose \(\mathbf a\) such that \(B\mathbf a=\Delta\Xvec\).
The translated field
\begin{equation}
\uvec_{\Xvec+\Delta\Xvec}^{\ast}(\rvec)
=
\uvec_{\Xvec}^{\ast}(\rvec+\mathbf a)
\label{eq:translated_solution}
\end{equation}
has the same bulk elastic and interfacial energy density after the change of variable $\mathbf r'=\mathbf r+\mathbf a$. Translation preserves the zero-mean constraint for periodic approximants and in the thermodynamic limit.
Hence
\begin{equation}
\lim_{A\to\infty}
\frac{
E_{\min}(\Xvec+\Delta\Xvec;A)
-
E_{\min}(\Xvec;A)
}{A}
=0
\label{eq:continuum_covariance}
\end{equation}
for boundary-condition sequences that restore translation invariance.
At the level of the infinite bulk functional, the translated textures form an exactly degenerate continuous path. For a finite sequence, the same conclusion requires the translation-induced boundary-energy difference and its derivative with respect to $\Xvec$ to be uniformly $o(A)$. The covariance maps the global minimum and any continuously translated stable branch within a fixed topological sector; it does not assert dynamical connectivity between disconnected sectors or remove barriers imposed by constraints absent from Eq.~\eqref{eq:energy_functional}. Under these conditions, smooth moir\'e reconstruction alone cannot establish an extensive equilibrium corrugation or an intrinsic bulk depinning stress. If $B$ is rank deficient, the construction applies only to shifts $\Delta\Xvec\in\operatorname{Im}B$; an unprotected stacking direction can remain a commensurate channel and must be tested separately.
Finite pinning requires a mechanism that breaks the relevant covariance, such as a commensurate channel, atomic discreteness, a Peierls barrier for domain-wall motion, edges, defects, or an externally imposed pinning field.
This is why a continuum reconstruction calculation and a static-friction phase calculation are not interchangeable.

\section{Elastic reconstruction parameter}

The harmonic elastic energy is written in Fourier space as
\begin{equation}
\frac{E_{\rm el}}{A}
=
\frac{1}{2}
\sum_{\qvec}
u_i(-\qvec)
D_{{\rm rel},ij}(\qvec)
u_j(\qvec).
\label{eq:Eelastic}
\end{equation}
We use $\uvec(\rvec)=\sum_{\qvec}\uvec(\qvec)e^{\ii\qvec\cdot\rvec}$; the sum runs over the reciprocal vectors of the chosen moir\'e cell or finite approximant, and $D_{{\rm rel},ij}$ is the kernel for relative in-plane displacement. Cartesian indices are summed when repeated.
For layer $\ell=1,2$ with in-plane elastic tensor $C_{ijkl}^{(\ell)}$,
\begin{equation}
D_{ik}^{(\ell)}(\qvec)
=C_{ijkl}^{(\ell)}q_jq_l.
\label{eq:anisotropic_kernel}
\end{equation}
If both layers relax elastically, the relative elastic kernel is
\begin{equation}
D_{\rm rel}(\qvec)
=
\left[
D_1^{-1}(\qvec)
+
D_2^{-1}(\qvec)
\right]^{-1}.
\label{eq:Drel}
\end{equation}
Here $D_1$ and $D_2$ are the individual layer kernels. Equation~\eqref{eq:Drel} follows by minimizing the sum of the two layer elastic energies at fixed relative displacement $\uvec=\uvec_1-\uvec_2$. If one layer is effectively rigid, \(D_{\rm rel}\) is the elastic kernel of the deformable layer.

Linear response to one moire harmonic gives
\begin{equation}
u_i(\qvec_n)
=
-\ii
\left[
D_{\rm rel}^{-1}(\qvec_n)
\right]_{ij}
G_{n,j}V_n+O(|V_n|^2).
\label{eq:linear_response}
\end{equation}
The dimensionless susceptibility to nonlinear registry feedback is therefore
\begin{equation}
\boxed{
\Lambda_n
=
|V_n|\,G_{n,i}
\left[
D_{\rm rel}^{-1}(\qvec_n)
\right]_{ij}
G_{n,j}.
}
\label{eq:lambda_tensor}
\end{equation}
Equation~\eqref{eq:lambda_tensor} applies to incommensurate harmonics with $\qvec_n\ne0$. An exactly commensurate channel has $\qvec_n=0$, for which this reconstruction susceptibility is singular and the barrier must be tested directly.
For compact reporting we use the largest single-harmonic susceptibility,
\begin{equation}
\Lambda=\max_n\Lambda_n.
\label{eq:lambda_max}
\end{equation}
This maximum is a screening statistic, not a complete reduction of a multiharmonic adhesion landscape: cooperative harmonic coupling is retained only in the full minimization. Relaxation is perturbative when the converged spectrum gives $\Lambda_n\ll1$ without a large cumulative multiharmonic response; any $\Lambda_n$ of order unity or larger warns that nonlinear reconstruction and soliton formation cannot be neglected.
A specified model may possess a reconstruction threshold
\begin{equation}
\Lambda=\Lambda_{\rm rec}({\cal S}),
\label{eq:lambda_boundary}
\end{equation}
where \({\cal S}\) denotes the complete model specification: lattice symmetry, harmonic content, elastic kernel, boundary conditions, and any allowed relaxation channels.
This is not, in general, the static pinning boundary.
Pinning still requires a nonzero thermodynamic \(\taudep\) after the microscopic covariance-breaking mechanism is included.

A critical pinning parameter becomes well defined only after a microscopic model family is specified. Let all stacking harmonics be scaled as \(V_n\mapsto gV_n\), with their ratios, lattice geometry, elastic tensor, boundary condition, defect content, and branch-selection rule collected in \({\cal S}\). For a monotone one-parameter pinning family and direction \(\hat{\mathbf e}\), define
\begin{equation}
\begin{aligned}
g_c(\hat{\mathbf e};{\cal S})
&=\inf\!\bigl\{g:\\[-2pt]
&\qquad\liminf_{A\to\infty}
\taudep(A,\hat{\mathbf e};g,{\cal S})>0\bigr\},
\end{aligned}
\label{eq:gc_definition}
\end{equation}
and
\begin{equation}
\Lambda_{\rm pin}(\hat{\mathbf e};{\cal S})
=g_c(\hat{\mathbf e};{\cal S})\Lambda(g=1).
\label{eq:lambda_pin_definition}
\end{equation}
Because this scaling gives $\Lambda(g)=g\Lambda(g=1)$, Eqs.~\eqref{eq:gc_definition} and \eqref{eq:lambda_pin_definition} define a critical parameter without assuming universality. Different atomic sampling, higher harmonics, or boundary ensembles may have the same bare \(\Lambda\) but different \(g_c\), which is why \(\Lambda_{\rm rec}\) and \(\Lambda_{\rm pin}\) must be distinguished. If the set in Eq.~\eqref{eq:gc_definition} is empty, we use the standard convention $g_c=\infty$.

The reconstruction susceptibility can also be obtained by nondimensionalizing the elastic relaxation problem.
For one dominant moire harmonic and one effective elastic channel, consider
\begin{equation}
E[\mathbf u]
=
\int \dd^2 r
\left[
\frac{C_{\rm eff}}{2}|\nabla\mathbf u|^2
+
2|V_G|\cos(\mathbf q\cdot\mathbf r+\mathbf G\cdot\mathbf u)
\right].
\label{eq:minimal_nondim_model}
\end{equation}
Here $q=|\mathbf q|$, $G=|\mathbf G|$, $C_{\rm eff}$ is the stiffness of the selected elastic channel, and the phase of the complex coefficient $V_G$ has been absorbed into the origin of stacking space. Let \(\boldsymbol\rho=q\mathbf r\), $\hat{\mathbf q}=\mathbf q/q$, and \(\phi=\mathbf G\cdot\mathbf u\).
Up to geometry-dependent numerical factors,
\begin{equation}
\begin{aligned}
E=\frac{C_{\rm eff}}{G^2}\int \dd^2\rho\,
\biggl[&\frac{1}{2}|\nabla_\rho\phi|^2\\
&+2\frac{|V_G|G^2}{C_{\rm eff}q^2}
\cos(\hat{\mathbf q}\cdot\boldsymbol\rho+\phi)\biggr].
\end{aligned}
\label{eq:nondim_model}
\end{equation}
Thus the nonlinear reconstruction problem is controlled by
\begin{equation}
\Lambda
\sim
\frac{|V_G|G^2}{C_{\rm eff}q^2}.
\label{eq:scalar_lambda_derivation}
\end{equation}
Equation~\eqref{eq:lambda_tensor} is the anisotropic and multi-harmonic generalization of this ratio.
The parameter is therefore not a dimensional guess: \(\Lambda\) is the dimensionless amplitude multiplying the nonlinear registry term after elastic rescaling.
Numerical prefactors enter a model-dependent reconstruction scale, while a pinning threshold, if present, must be obtained from the displacement-dependent relaxed energy.

The familiar one-dimensional Pokrovsky--Talapov reference illustrates why its Fourier convention must be stated. For the one-dimensional reduction of Eq.~\eqref{eq:minimal_nondim_model}, the full cosine amplitude is $2|V_G|$. An isolated $2\pi$ soliton costs, per unit transverse length,
\begin{equation}
\begin{aligned}
E_{\rm sol}&=\frac{8}{G}\sqrt{2C_{\rm eff}|V_G|},\\
\Delta E_{\rm mis}&=\frac{2\pi C_{\rm eff}q}{G^2},
\end{aligned}
\label{eq:pt_balance}
\end{equation}
where $\Delta E_{\rm mis}$ is the energy gained from the imposed mismatch. Equating the two gives
\begin{equation}
\Lambda_{\rm PT}
=\frac{|V_G|G^2}{C_{\rm eff}q_c^2}
=\frac{\pi^2}{32}.
\label{eq:pt_complex_convention}
\end{equation}
Here $q_c$ is the critical one-dimensional mismatch. If the coupling is instead defined with the full cosine amplitude $U=2|V_G|$, the same result reads $UG^2/(C_{\rm eff}q_c^2)=\pi^2/16$ \cite{PokrovskyTalapov1979}. This is a one-dimensional reconstruction threshold, not an Aubry or static-pinning threshold.

The length-scale interpretation is equivalent.
The moir\'e length is of order \(\ell_M\sim q^{-1}\), while the soliton width is of order
\begin{equation}
\ell_s
\sim
\frac{1}{G}
\sqrt{\frac{C_{\rm eff}}{|V_G|}}.
\label{eq:ell_s}
\end{equation}
Therefore
\begin{equation}
\Lambda
\sim
\left(
\frac{\ell_M}{\ell_s}
\right)^2.
\label{eq:length_ratio}
\end{equation}
If \(\ell_M\gg\ell_s\), the interface can form commensurate domains separated by narrow solitons.
If \(\ell_M\ll\ell_s\), reconstruction is perturbative.
Neither limit alone fixes the thermodynamic static-friction phase.

\section{Two-dimensional anisotropy and the hexagonal limit}

Equation~\eqref{eq:lambda_tensor} is the primary two-dimensional result.
It does not require isotropic elasticity.
For a low-symmetry material, \(D_{ij}(\qvec)\) must be built from the full two-dimensional elastic tensor.
The orientation dependence of \(D_{ij}\) changes the preferred domain-wall directions and the twist or mismatch at which strong reconstruction sets in.

For hexagonal monolayers at long wavelength, the quadratic in-plane elastic kernel is isotropic:
\begin{equation}
D_{ij}(\qvec)
=
C_L q^2 \hat q_i\hat q_j
+
C_T q^2
(\delta_{ij}-\hat q_i\hat q_j),
\label{eq:isotropic_D}
\end{equation}
where $\hat q_i=q_i/q$, $\delta_{ij}$ is the Kronecker delta, \(C_L=C_{11}\), and \(C_T=C_{66}=(C_{11}-C_{12})/2\).
For two relaxing layers,
\begin{equation}
\begin{aligned}
\frac{1}{C_L^{\rm rel}}
&=\frac{1}{C_L^{(1)}}+\frac{1}{C_L^{(2)}},\\
\frac{1}{C_T^{\rm rel}}
&=\frac{1}{C_T^{(1)}}+\frac{1}{C_T^{(2)}}.
\end{aligned}
\label{eq:relative_CL_CT}
\end{equation}
Substitution into Eq.~\eqref{eq:lambda_tensor} gives
\begin{equation}
\Lambda_n
=
\frac{|V_n|}{q_n^2}
\left[
\frac{G_{n,L}^2}{C_L^{\rm rel}}
+
\frac{G_{n,T}^2}{C_T^{\rm rel}}
\right],
\label{eq:lambda_LT}
\end{equation}
where \(G_{n,L}\) and \(G_{n,T}\) are the components of \(\Gvec_n\) parallel and perpendicular to \(\qvec_n\).

For two nearly aligned hexagonal lattices, define \(\delta\) as the small reciprocal-lattice mismatch normalized to $G_n$ and \(\theta\) as the twist angle in radians.
To leading order,
\begin{equation}
\qvec_n
\simeq
\delta\,\Gvec_n
-
\theta\,\hat{\mathbf z}\times\Gvec_n.
\label{eq:q_delta_theta}
\end{equation}
Therefore
\begin{equation}
q_n^2
=
G_n^2(\delta^2+\theta^2),
\label{eq:q_square}
\end{equation}
and
\begin{equation}
\begin{aligned}
G_{n,L}^2
&=G_n^2\frac{\delta^2}{\delta^2+\theta^2},\\
G_{n,T}^2
&=G_n^2\frac{\theta^2}{\delta^2+\theta^2}.
\end{aligned}
\label{eq:GL_GT}
\end{equation}
For a first-star isotropic corrugation \(V_n=V_G\), the reconstruction susceptibility becomes
\begin{equation}
\boxed{
\Lambda(\theta,\delta)
=
|V_G|
\frac{
\delta^2/C_L^{\rm rel}
+
\theta^2/C_T^{\rm rel}
}{
(\delta^2+\theta^2)^2
}.
}
\label{eq:lambda_hex}
\end{equation}
Equivalently,
\begin{equation}
\Lambda(\theta,\delta)
=
\frac{|V_G|}
{C_{\rm eff}(\theta,\delta)(\delta^2+\theta^2)},
\label{eq:lambda_Ceff}
\end{equation}
with
\begin{equation}
\frac{1}{C_{\rm eff}(\theta,\delta)}
=
\frac{
\delta^2/C_L^{\rm rel}
+
\theta^2/C_T^{\rm rel}
}{
\delta^2+\theta^2
}.
\label{eq:Ceff}
\end{equation}

Equations~\eqref{eq:superlubric_definition} and \eqref{eq:pinned_definition} define the thermodynamic static-friction phases.
The reduced coordinates $x=q/G$ and $y=|V_G|/C_{\rm eff}$ give $\Lambda=y/x^2$, but an $(x,y)$ diagram is only a reconstruction map.
For \(\Lambda\ll1\), elastic relaxation is perturbative; when \(\Lambda\) is of order unity or larger, domain formation can become energetically favorable.
The Pokrovsky--Talapov benchmark and the elastic crossover are therefore guides to reconstruction, not universal pinning boundaries.
A point can be assigned to the pinned, direction-selective, or fully superlubric phase only after evaluating \(\taudep(A,\hat{\mathbf e})\), supported by \(\taueq(A)\) and \(\Phi_R(A)\), with the relevant atomistic, boundary, and defect physics.

\section{Graphene/hBN adhesion and elastic input}

Graphene/hBN is a stringent test because its small mismatch favors moir\'e relaxation, while low-friction graphite/hBN contacts have also been observed \cite{Woods2014,Song2018}. Recent DMC calculations determine the adhesion energy as a function of local offset and report no metastable relaxed structure within their continuum model \cite{Szyniszewski2025}. We use those coefficients without fitting to friction.

The DMC geometry is
\begin{equation}
a_{\rm G}=2.462~{\rm \Ang},\qquad
a_{\rm hBN}=2.504~{\rm \Ang},
\end{equation}
which gives the reciprocal mismatch relevant to $q/G$,
\begin{equation}
\begin{aligned}
\delta&=1-\frac{a_{\rm G}}{a_{\rm hBN}}=0.0167732,\\
\Acell&=\frac{\sqrt3}{2}a_{\rm G}^2=5.24936~{\rm \Ang^2}.
\end{aligned}
\label{eq:ghbn_geometry_qmc}
\end{equation}
The corresponding zero-twist moir\'e length is approximately $14.7~{\rm nm}$.

The DMC adhesion potential per graphene primitive cell is written as \cite{Szyniszewski2025}
\begin{equation}
V_A(\bm\ell)=v_{s0}
+v_{s1}\sum_{j=1}^{3}\cos(\mathbf G_j\cdot\bm\ell)
+v_{as1}\sum_{j=1}^{3}\sin(\mathbf G_j\cdot\bm\ell),
\label{eq:dmc_adhesion}
\end{equation}
where $\bm\ell$ is the local in-plane stacking displacement, $\mathbf G_j$ are the three positive first-star reciprocal vectors, and $v_{s0}$ is a stacking-independent offset that does not affect reconstruction or friction. The fitted corrugation coefficients are
\begin{equation}
v_{s1}=2.2(3),\qquad v_{as1}=-3.5(4)
\quad {\rm meV/cell}.
\end{equation}
Parentheses denote the reported one-standard-deviation statistical uncertainty in the final digit. Matching Eq.~\eqref{eq:dmc_adhesion} to Eq.~\eqref{eq:Vint}, and writing the per-cell coefficient as $E_G=\Acell V_G$, gives
\begin{equation}
E_G=\frac{v_{s1}-\ii v_{as1}}{2}
=(1.10+\ii1.75)~{\rm meV/cell},
\end{equation}
and therefore
\begin{equation}
|E_G|=2.067~{\rm meV/cell},\qquad
|V_G|=3.9376\times10^{-4}~{\rm eV/\Ang^2}.
\label{eq:dmc_fourier_result}
\end{equation}
This conversion is essential: neither the peak-to-peak barrier nor the $15~{\rm meV/cell}$ mean-to-minimum adhesion reduction reported for a uniformly lattice-matched structure is the single complex coefficient $E_G$.

For graphene we use $C_{11}=352.6$, $C_{12}=59.6$, and $C_{66}=146.5~{\rm N/m}$ \cite{Liu2016Elastic}; for hBN we use $C_{11}=293.2$, $C_{12}=66.1$, and $C_{66}=113.55~{\rm N/m}$ \cite{Peng2012}. With $1~{\rm N/m}=0.0624151~{\rm eV/\Ang^2}$, Eq.~\eqref{eq:relative_CL_CT} yields
\begin{equation}
C_L^{\rm rel}=9.9917~{\rm eV/\Ang^2},\qquad
C_T^{\rm rel}=3.9926~{\rm eV/\Ang^2}
\label{eq:ghbn_relative_moduli}
\end{equation}
for two free layers. For rigid hBN, the graphene values are $C_L=22.0076$ and $C_T=9.14381~{\rm eV/\Ang^2}$.

At zero twist, the DMC input gives
\begin{align}
\Lambda_0^{\rm free/free}&=0.1401,\nonumber\\[-2pt]
{\cal I}_{95}^{\rm free/free}&=[0.1158,0.1654],\\
\Lambda_0^{\rm rigid\ hBN}&=0.06360,\nonumber\\[-2pt]
{\cal I}_{95}^{\rm rigid\ hBN}&=[0.05258,0.07509].
\label{eq:ghbn_lambda_results}
\end{align}
Here ${\cal I}_{95}$ denotes the 95\% DMC-statistical interval. The intervals propagate only the published statistical errors of $v_{s1}$ and $v_{as1}$; they do not include elastic or model-form uncertainty. Formally solving $g\Lambda_0=1$ gives $g=7.14$ for free/free and $15.72$ for rigid hBN. These numbers are reconstruction scales, not pinning critical points.

The weak reconstruction inferred from the DMC first star is not equivalent to a claim that every experimental graphene/hBN specimen is weakly reconstructed. Domain formation reported near alignment \cite{Woods2014} can depend on higher adhesion harmonics, multilayer boundary conditions, out-of-plane relaxation, residual strain, and preparation. We test the first source of model-form uncertainty with the Leven potential below; out-of-plane and nonlinear intralayer relaxation remain separate limitations. The independent QMC-fitted interlayer potential of Krongchon \emph{et al.} provides a natural next test of model form and height relaxation \cite{Krongchon2025}.
Table~\ref{tab:ghbn_qmc_input} collects the published elastic and DMC inputs and the resulting zero-twist susceptibilities.

\begin{table}[t]
\caption{Published input and derived zero-twist quantities for graphene/hBN. The uncertainty intervals propagate only the DMC Fourier-fit errors.}
\label{tab:ghbn_qmc_input}
\begingroup
\footnotesize
\setlength{\tabcolsep}{3pt}
\begin{ruledtabular}
\begin{tabular}{lcc}
Quantity & free/free & rigid hBN \\
\hline
$C_L^{\rm rel}$ ($\mathrm{eV/\Ang^2}$) & 9.9917 & 22.0076 \\
$C_T^{\rm rel}$ ($\mathrm{eV/\Ang^2}$) & 3.9926 & 9.14381 \\
$\Lambda_0$ & 0.1401 & 0.06360 \\
DMC 95\% interval & $[0.1158,0.1654]$ & $[0.05258,0.07509]$ \\
$g$ at $\Lambda=1$ & 7.14 & 15.72 \\
\end{tabular}
\end{ruledtabular}
\endgroup
\end{table}

\emph{Independent ILP audit.---}
We evaluate the C--B and C--N pair form and Table-I parameters of Leven \emph{et al.} at the fixed $3.3~{\rm \Ang}$ separation used for their published sliding landscape \cite{Leven2016}. The local commensurate grid uses $a_0=2.504~{\rm \Ang}$ and area $A_0=\sqrt{3}a_0^2/2=5.42999~{\rm \Ang^2}$. Its $48\times48$ peak-to-trough corrugation is $30.420~{\rm meV}$ per local cell, or $5.602~{\rm meV/\Ang^2}$ when divided by $A_0$. Retaining 15 reciprocal pairs down to $10^{-5}$ of the largest Fourier amplitude reproduces this grid with a root-mean-square error of $3.3\times10^{-5}~{\rm meV/cell}$. The largest complex coefficient is $3.0857~{\rm meV/cell}$, compared with $2.067~{\rm meV/cell}$ for the DMC first star. On transfer to the common discrete graphene sampling model, we retain these per-cell energies and divide by the graphene area $\Acell$ when forming $V_n$; this normalization matches Eq.~\eqref{eq:computed_dfk} and is recorded in the ancillary data. Defining $\Lambda_{\max}^{\rm test}$ as the largest single-harmonic susceptibility over the three approximants, we obtain $0.2118$ for free/free and $0.09616$ for rigid hBN. Thus the ILP is the more reconstructive of the two tested landscapes, while the relaxed discrete solutions at physical coupling remain in the small-strain regime.

\section{Two-dimensional discrete pinning test}

To test whether atomic sampling converts reconstruction into pinning, we use a triangular-lattice discrete FK model with a general Fourier-resolved adhesion spectrum,
\begin{align}
E_{\rm dFK}[\mathbf u;\mathbf X]
={}&\frac{A}{2}\sum_{\mathbf q\ne0}
u_i^*(\mathbf q)D_{{\rm rel},ij}(\mathbf q)u_j(\mathbf q)\nonumber\\
&+g\sum_{\mathbf R}\sum_{\alpha\in\mathcal H}
\left[\mathcal C_\alpha\cos\phi_{\alpha\mathbf R}
+\mathcal S_\alpha\sin\phi_{\alpha\mathbf R}\right],
\label{eq:computed_dfk}
\end{align}
where
\begin{equation}
\phi_{\alpha\mathbf R}=\varphi_{\alpha\mathbf R}
+\mathbf G_\alpha\cdot
[\mathbf X+\mathbf u(\mathbf R)],
\qquad \sum_{\mathbf R}\mathbf u(\mathbf R)=0.
\end{equation}
Here $\mathbf R$ runs over the $N^2$ sites of a triangular graphene supercell, $A=N^2\Acell$ is its area, $\mathcal H$ is the retained set of adhesion harmonics, and $\varphi_{\alpha\mathbf R}=\mathbf G_\alpha\cdot B\mathbf R$ modulo $2\pi$ is the unrelaxed mismatch phase. The real coefficients $\mathcal C_\alpha$ and $\mathcal S_\alpha$ are adhesion energies assigned per graphene sampling site. For DMC, $\mathcal H$ contains the three first-star vectors and $(\mathcal C_\alpha,\mathcal S_\alpha)=(v_{s1},v_{as1})$; for the Leven model it contains the 15 retained reciprocal pairs obtained from the fixed-height stacking grid. The dimensionless factor $g$ scales the entire adhesion spectrum: physical coupling is $g=1$, while $g>1$ is used only for the DMC sensitivity scan. The elastic kernel is Eq.~\eqref{eq:isotropic_D}.
At zero twist we index $\mathbf G_\alpha=h\mathbf b_1^{\rm S}+k\mathbf b_2^{\rm S}$ on the substrate reciprocal basis, dual to the real-space vectors $\mathbf a_1^{\rm S},\mathbf a_2^{\rm S}$.

We approximate the target zero-twist mismatch $\delta_\ast=\delta=0.0167732$ by the coprime ratios $\delta_N=p/N=1/60$, $2/119$, and $3/179$, with $3600$, $14161$, and $32041$ displacement sites. Their supercell side lengths are $14.772$, $29.2978$, and $44.0698~{\rm nm}$, while the corresponding moir\'e periods remain $14.772$, $14.6489$, and $14.6899~{\rm nm}$. The incommensurate thermodynamic limit is taken along a coprime sequence $p_k/N_k\to\delta_\ast$ with $A_k=N_k^2\Acell\to\infty$. Periodically repeating one fixed rational cell does not approach this limit: it defines a different, exactly commensurate model and can retain an artificial extensive barrier. Let $\mathbf h=h_1\mathbf a_1^{\rm S}+h_2\mathbf a_2^{\rm S}$ be a primitive substrate lattice vector. For a coprime $p/N$ approximant the relaxed Peierls landscape along $\mathbf h$ has period $|\mathbf h|/N$; sampling a full substrate period at divisors of $N$ would alias the barrier. We sample 24 shifts over this primitive period and five coprime directions spanning the irreducible $0^\circ$--$30^\circ$ sector. The $3/179$ approximant is also retained in the strongest-coupling size audit.

Each constrained minimum is obtained by spectral-preconditioned L-BFGS followed by damped Newton--conjugate-gradient refinement with an analytic Hessian. The analytic gradient agrees with centered finite differences to a maximum relative error of $7.23\times10^{-10}$; translation by one substrate vector and the $g=0$ limit are satisfied to machine precision. Fourier normalization, convergence tests, complete pathwise tables, and the covariance derivation are given in the Supplemental Material \cite{SupplementalMaterial}. Along each sampled path, $\tau_{\rm force}=A_N^{-1}\hat{\mathbf e}\cdot\partial E/\partial\mathbf X$ is the direct generalized stress, whereas $\tau_{\rm energy}$ is obtained by differentiating the Fourier representation of $E_{\min}/A_N$. The physical restoring stress has the opposite sign; all reported diagnostics use absolute values or squared differences. With $\langle\cdots\rangle$ denoting the average over the 24 shifts on that path, we define the conservative numerical resolution
\begin{equation}
\begin{aligned}
\epsilon_\tau=\max\bigl[&
5\langle(\tau_{\rm force}-\tau_{\rm energy})^2\rangle^{1/2},\\
&10^{-10}~{\rm MPa}\bigr].
\end{aligned}
\label{eq:numerical_resolution}
\end{equation}
The factor five is a safety multiplier on the root-mean-square disagreement of two independent stress estimators, and $10^{-10}~{\rm MPa}$ prevents machine-precision cancellation from producing a vanishing threshold. A barrier is reported only if it exceeds $\epsilon_\tau$.

At $g=1$ no direction in any of the three approximants has a resolved equilibrium barrier. For the sampled set $\mathcal D_{\rm samp}$ of five directions, define
\begin{equation}
\tau_{\rm eq}^{\max,{\rm samp}}(N)
=\max_{\mathbf h\in\mathcal D_{\rm samp}}\tau_{\rm eq}(A_N,\hat{\mathbf h}).
\end{equation}
Taking the largest pathwise resolution over all three sizes and all sampled directions gives
\begin{equation}
\begin{aligned}
\max_N\tau_{\rm eq}^{\max,{\rm samp}}(N)
&<2.7\times10^{-9}~{\rm MPa} &&(\mathrm{free/free}),\\
\max_N\tau_{\rm eq}^{\max,{\rm samp}}(N)
&<1.7\times10^{-8}~{\rm MPa} &&(\mathrm{rigid\ hBN}).
\end{aligned}
\label{eq:ghbn_stress_bounds}
\end{equation}
The largest displacement magnitudes are $0.1696$ and $0.0779~{\rm \Ang}$, and the largest strain components are $0.784\%$ and $0.338\%$, for free/free and rigid-hBN boundaries, respectively. Thus the physical DMC calculations remain within the small-strain regime.

To remove the one-dimensional-path ambiguity from the equilibrium diagnostic, we additionally minimize on a $12\times12$ grid spanning the primitive two-dimensional shift torus for every size and boundary condition. Let $\boldsymbol\tau_{\rm force}$ and $\boldsymbol\tau_{\rm energy}$ denote the direct and spectral two-component stresses on this grid. For each $N$ we use
\begin{equation}
\begin{aligned}
\epsilon_{\rm torus}(N)=\max\!\Bigl[
&5\bigl\langle
|\boldsymbol\tau_{\rm force}-\boldsymbol\tau_{\rm energy}|^2
\bigr\rangle_{\Omega_X}^{1/2},\\
&10^{-10}~{\rm MPa}\Bigr],
\end{aligned}
\label{eq:torus_resolution}
\end{equation}
and, separately for each boundary condition, define $\epsilon_{\rm torus}^{\max}=\max_N\epsilon_{\rm torus}(N)$. Spectral differentiation of $E_{\min}(\mathbf X)/A_N$ gives the energy-derived diagnostic $\Phi_R^{\rm energy}$ and the bounds
\begin{equation}
\begin{aligned}
\max_N\Phi_R^{\rm energy}
&=1.12\times10^{-11}~{\rm MPa}\\[-2pt]
&<3.41\times10^{-9}~{\rm MPa}
=\epsilon_{\rm torus}^{\max},\quad(\mathrm{free/free}),\\
\max_N\Phi_R^{\rm energy}
&=7.73\times10^{-12}~{\rm MPa}\\[-2pt]
&<9.24\times10^{-10}~{\rm MPa}
=\epsilon_{\rm torus}^{\max},\quad(\mathrm{rigid\ hBN}),
\end{aligned}
\label{eq:ghbn_torus_bounds}
\end{equation}
for free/free and rigid-hBN boundaries, respectively. The largest spectral restoring stresses are also below these independently defined force--energy consistency scales. Thus the full sampled torus contains no resolved coherent equilibrium corrugation; the quoted $\Phi_R$ values are numerical estimates below resolution, not positive order parameters.

We also minimize from 25 initial fields for each boundary condition, including random displacements with standard deviations up to $1~{\rm \Ang}$. All 50 endpoints collapse to the same configuration within $8\times10^{-10}~{\rm \Ang}$ and an energy spread below $8\times10^{-16}~{\rm meV/cell}$. Forty-eight endpoints satisfy the strict residual-force threshold; the other two coincide with the same solution but stop above that numerical threshold. No distinct zero-stress metastable branch is found.

For a fixed approximant of area $A_N$, we explicitly continue each observed branch along a sampled direction and write $\mathbf X=X\hat{\mathbf e}$, where $X$ is the scalar unwrapped coordinate and $E_m(X)\equiv E_m(X\hat{\mathbf e};A_N)$. The reduced tilted enthalpy is
\begin{equation}
h_{m,\tau}(X)=\frac{E_m(X)}{A_N}-\tau X.
\label{eq:reduced_tilted_enthalpy}
\end{equation}
Because the internal coordinates are minimized at fixed $X$, the effective branch curvature is the Schur complement
\begin{equation}
\kappa_{\rm eff}
=E_{XX}-E_{Xu}H_{uu}^{-1}E_{uX}.
\label{eq:schur_curvature}
\end{equation}
Here $E_{XX}$ is the second derivative of the full energy with respect to $X$, $E_{Xu}$ and $E_{uX}$ are the mixed derivative blocks, and $H_{uu}$ is the Hessian with respect to the mean-zero internal displacement coordinates, all evaluated on the branch. A stable stationary point satisfies $\partial_XE_m/A_N=\tau$ and $\kappa_{\rm eff}>0$; depinning occurs at its saddle-node, where $\kappa_{\rm eff}=0$. Define $\tau_{\rm dep}^{\max,{\rm samp}}(N)=\max_{\mathbf h\in\mathcal D_{\rm samp}}\taudep(A_N,\hat{\mathbf h})$. At physical $g=1$, no finite-stress saddle-node or terminating metastable branch is resolved above the force--energy consistency scale in any sampled direction, giving
\begin{equation}
\begin{aligned}
\max_N\tau_{\rm dep}^{\max,{\rm samp}}(N)
&<2.7\times10^{-9}~{\rm MPa} &&(\mathrm{free/free}),\\
\max_N\tau_{\rm dep}^{\max,{\rm samp}}(N)
&<1.7\times10^{-8}~{\rm MPa} &&(\mathrm{rigid\ hBN}).
\end{aligned}
\label{eq:ghbn_depinning_bounds}
\end{equation}
These are null-detection bounds for the observed branches, not positive depinning stresses or a proof that a remote unsampled branch cannot exist.

Repeating all 30 physical-coupling paths with the 15-harmonic Leven spectrum gives no resolved barrier. The largest conservative force--energy resolutions are $3.2\times10^{-9}~{\rm MPa}$ for free/free and $4.7\times10^{-10}~{\rm MPa}$ for rigid hBN. The largest local strain components are $1.17\%$ and $0.516\%$, respectively, so this comparison does not rely on the uncontrolled strong-strain regime.

The $12\times12$ two-dimensional shift-torus and 25-start searches were then repeated with the Leven spectrum. All six torus scans converge. For free/free boundaries the largest $\Phi_R$ and maximum equilibrium stress are $4.57\times10^{-11}$ and $1.24\times10^{-10}~{\rm MPa}$, below the largest consistency scale $2.04\times10^{-9}~{\rm MPa}$; for rigid hBN the corresponding values are $1.90\times10^{-11}$ and $5.16\times10^{-11}~{\rm MPa}$, below $1.27\times10^{-8}~{\rm MPa}$. All 50 multi-start runs converge to the same zero-stress state within the optimizer tolerance. Table~\ref{tab:ghbn_model_robustness} separates the increase in reconstruction strength from the unchanged null result; complete torus and multi-start data are given in the Supplemental Material.

\begin{table}[t]
\caption{Adhesion-model robustness over three approximants and five directions. Bounds are the largest pathwise numerical resolutions; no row contains a resolved nonzero barrier.}
\label{tab:ghbn_model_robustness}
\begingroup
\footnotesize
\setlength{\tabcolsep}{3pt}
\begin{ruledtabular}
\begin{tabular}{lccc}
Model/boundary & $\Lambda_{\max}^{\rm test}$ & max. strain & bound (MPa) \\
\hline
DMC/free & 0.1419 & 0.784\% & $2.7\times10^{-9}$ \\
DMC/rigid & 0.06441 & 0.338\% & $1.7\times10^{-8}$ \\
Leven/free & 0.2118 & 1.17\% & $3.2\times10^{-9}$ \\
Leven/rigid & 0.09616 & 0.516\% & $4.7\times10^{-10}$ \\
\end{tabular}
\end{ruledtabular}
\endgroup
\end{table}

To search for a pinning point, all DMC stacking harmonics are scaled by $g$. The smallest approximant develops resolved saddle-nodes at $g=40$ and $80$, with $\tau_{\rm dep}=1.43\times10^{-6}$ and $2.13\times10^{-3}~{\rm MPa}$. These finite-approximant thresholds do not survive the size test. At $g=80$, the energy corrugation per cell falls by a factor $4.76\times10^3$ from $N=60$ to $N=119$ and by a further factor $6.62$ from $N=119$ to $N=179$; the latter two branches lie below their force--energy resolution. The maximum local strain is already $9.5\%$, $14.2\%$, and $20.7\%$ at $g=20$, $40$, and $80$. Harmonic elasticity and a fixed first-star adhesion potential are not controlled at those strains.
Complete strong-coupling values for every approximant are given in the Supplemental Material.

Figure~\ref{fig:ghbn_results} collects the reconstruction screen, the physical-coupling size flow, and the artificial strong-coupling stability tests.
\begin{figure*}[t]
\centering
\includegraphics[width=0.98\textwidth]{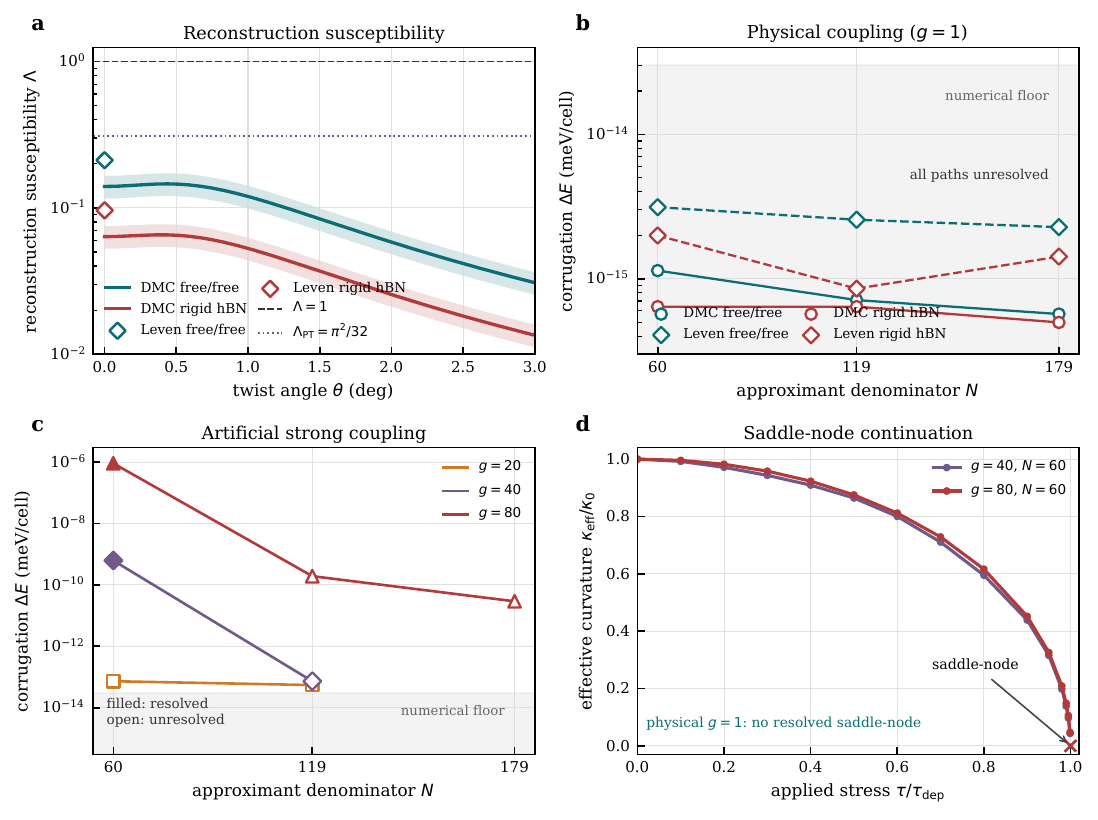}
\caption{
Reconstruction and depinning tests for graphene/hBN.
(a) Twist-dependent $\Lambda$ obtained from the DMC first-star adhesion surface; shaded bands propagate the reported Fourier-fit uncertainties. Diamonds mark the largest zero-twist values found among the tested Leven-spectrum approximants. The dashed line marks the reconstruction scale $\Lambda=1$, and the dotted line is the one-dimensional Pokrovsky--Talapov value in the complex-coefficient convention of Eq.~\eqref{eq:pt_complex_convention}; neither line is a pinning boundary.
(b) Physical-coupling ($g=1$) finite-size flow of the largest constrained corrugation among five loading directions for both adhesion models and boundary conditions. All symbols are open because every path lies below the a priori specified force--energy resolution.
(c) The same finite-size test after artificially scaling the DMC adhesion by $g=20$, $40$, and $80$. Filled symbols are resolved and open symbols are unresolved; the signals first seen at $N=60$ do not persist at the larger tested approximants.
(d) Tilted-enthalpy continuation of the two resolved $N=60$ branches. Here $\kappa_0$ is the zero-stress branch curvature used to normalize $\kappa_{\rm eff}$; the normalized curvature vanishes at a saddle-node. No physical-$g=1$ branch is resolved.
}
\label{fig:ghbn_results}
\end{figure*}

No controlled critical coupling emerges from this scan. A size-independent pinning barrier is not resolved before the model leaves its small-strain regime, so $g_c>80$ would not be a defensible material bound. The physical DMC point $g=1$ lies in the weak-reconstruction, unresolved-barrier region, and crossing $\Lambda=1$ does not mark the static transition.

\section{Finite size and polycrystallinity}

The phase criterion is based on scaling with area, but finite contacts contain geometrically distinct force contributions. In this section the loading direction and the sequence of shapes, boundary conditions, and preparation protocols are held fixed, and the directional argument is suppressed. Let $F_{\rm dep}(A)=A\taudep(A)$ be the directional depinning force. Its model-independent decomposition is
\begin{equation}
F_{\rm dep}(A)=\taudep^{\rm bulk}A+F_{\rm sub}(A),
\qquad
\lim_{A\to\infty}\frac{F_{\rm sub}(A)}{A}=0.
\label{eq:general_finite_size}
\end{equation}
The function $F_{\rm sub}$ depends on shape, edge orientation, and boundary conditions; no single power law describes every finite slider. For an edge-dominated sequence of fixed-shape polygonal contacts, a useful envelope is
\begin{align}
F_{\rm dep}(A)
&=a_{\rm c}+b_{\rm e}A^{1/2}+\taudep^{\rm bulk}A
+o(A^{1/2}),\nonumber\\
\taudep(A)
&=\taudep^{\rm bulk}+\frac{b_{\rm e}}{A^{1/2}}
+\frac{a_{\rm c}}{A}+o(A^{-1/2}),
\label{eq:edge_scaling}
\end{align}
Here $\taudep^{\rm bulk}$ is a stress, $a_{\rm c}$ is the net bounded corner contribution with units of force, and $b_{\rm e}$ is an edge-envelope coefficient with units of force per length; $o(A^{1/2})$ denotes terms asymptotically smaller than $A^{1/2}$. The calculations of Yan and Gao show why Eq.~\eqref{eq:general_finite_size}, rather than a universal perimeter fit, is the appropriate starting point: polygonal edge envelopes can scale as $A^{1/2}$, rigid circular flakes can show $A^{1/4}$ behavior, and generic polygons carry bounded moir\'e oscillations \cite{YanGao2024Shape,Gao2025Edge}. Each contribution is subextensive. The phase is set by the large-area intercept $\taudep^{\rm bulk}$ after the appropriate shape-dependent terms are removed. If the $A^{1/2}$ edge term dominates the finite-size correction and $\taudep^{\rm bulk}>0$, the edge-to-bulk crossover is $A_\times=(b_{\rm e}/\taudep^{\rm bulk})^2$.

Gao \emph{et al.} also proposed the kinetic law $F_k=a_k+b_kA^{1/2}+c_kA$ \cite{Gao2025Edge}, where $a_k$, $b_k$, and $c_k$ are the kinetic corner, edge, and area coefficients, with units of force, force per length, and stress, respectively. Its algebraic separation is useful experimentally, but $c_k$ measures dissipative surface friction and is not the static order parameter $\taudep^{\rm bulk}$. Static and kinetic extrapolations must therefore not be interchanged. Finite static friction in a small flake is not sufficient evidence for a thermodynamic pinned phase.

For a polycrystalline interface, decompose the total resisting force into grain-interior, grain-boundary, edge, and point-defect contributions,
\begin{equation}
\mathbf F_{\rm dep}^{\rm poly}
=
\sum_\alpha\mathbf F_\alpha
+\mathbf F_{\rm GB}
+\mathbf F_{\rm edge}
+\mathbf F_{\rm def}.
\label{eq:poly}
\end{equation}
Here $\alpha$ labels grains, $\mathbf F_\alpha$ is the resisting force from grain $\alpha$, and $\mathbf F_{\rm GB}$, $\mathbf F_{\rm edge}$, and $\mathbf F_{\rm def}$ are the grain-boundary, outer-edge, and point-defect contributions. Random grain registry phases make the coherent grain-interior sum subextensive when its correlation length remains finite. By contrast, a connected grain-boundary network or a finite areal density of strongly pinned defects can produce an extensive contribution.
Atomistic simulations show that grain-boundary corrugation and buckling can dominate dissipation in polycrystalline graphene \cite{Gao2021}. Finite single-crystal contacts likewise require explicit edge scaling, as established in atomistic theory and direct graphite measurements \cite{Varini2015,Gigli2017,Hu2024,Li2024ZeroStatic}.
This is why single-crystal and polycrystalline contacts require the same force-per-area limit but generally belong to different finite-size scaling ensembles.

\section{Implications for materials screening}

The criterion gives a compact screening protocol:
\begin{equation}
\begin{aligned}
\{B,\Gvec_n,V_n,C_{ijkl}^{(\ell)}\}
&\rightarrow
\{\qvec_n,D_{\rm rel}(\qvec_n),\Lambda_n\},\\
&\rightarrow E_{\min}(\Xvec;A_k),\\
&\rightarrow\{\Phi_R(A_k),\taudep(A_k,\hat{\mathbf e})\},\\
&\rightarrow {\rm phase}.
\end{aligned}
\label{eq:screening}
\end{equation}
Here $B$ is the lattice-mismatch tensor, $C_{ijkl}^{(\ell)}$ is the in-plane elastic tensor of layer $\ell=1,2$, and $\{A_k\}$ is a specified increasing-area sequence with fixed shape, boundary condition, and branch-selection protocol. Large moir\'e wave vectors, weak corrugation, and high stiffness suppress reconstruction. Small mismatch or twist, strong corrugation, and low stiffness favor it. These trends make $\Lambda_n$ an inexpensive first-stage filter, but only the second-stage finite-size calculation assigns a static phase. Experiments on large-mismatch MoS\(_2\)/graphite and MoS\(_2\)/hBN interfaces are consistent with this screening logic while also showing that edge pinning remains measurable \cite{Liao2022}. First-principles indices and machine-learning models can efficiently rank candidates by stiffness, binding, and finite-cell sliding barriers \cite{Gao2021Screen,Chen2026ML}; those quantities remain screening descriptors rather than thermodynamic phase labels until the constrained barrier and depinning stress are extrapolated with size.
Table~\ref{tab:materials_screening} states what must be calculated before the same logic can be applied to other interface classes.

\begin{table}[t]
\caption{Candidate classes and the calculations needed beyond the first-stage reconstruction screen. None of the entries is a phase assignment.}
\label{tab:materials_screening}
\begingroup
\footnotesize
\setlength{\tabcolsep}{3pt}
\begin{tabular}{p{0.22\columnwidth}p{0.24\columnwidth}p{0.34\columnwidth}}
\hline
Interface class & First-stage expectation & Decisive second-stage test \\
\hline
graphene/hBN & small mismatch; $\Lambda_{\max}^{\rm test}<0.22$ here & height relaxation, nonlinear elasticity, dense angular scaling \\
graphene or hBN/TMD & mismatch suppresses $\Lambda$ if $|V_G|/C_{\rm eff}$ is comparable & load-dependent GSFE, edge scaling, and defects \\
near-matched TMD pairs & strong reconstruction is plausible & multistar GSFE and discrete finite-size flow \\
anisotropic interfaces & $\Lambda_n$ depends on direction & full $D_{ij}(\mathbf q)$ and directional depinning \\
\hline
\end{tabular}
\endgroup
\end{table}

The geometric effect of mismatch can be isolated without turning it into a material prediction. Writing \(x=q/G\), two interfaces obey the scaling estimate
\begin{equation}
\frac{\Lambda_2}{\Lambda_1}
\simeq
\frac{(|V_G|/C_{\rm eff})_2}{(|V_G|/C_{\rm eff})_1}
\left(
\frac{x_1}{x_2}
\right)^2.
\label{eq:lambda_ratio}
\end{equation}
For graphene/hBN, \(q/G\simeq\delta\simeq0.0168\).
For graphene/MoS\(_2\) \cite{Liao2022},
\begin{equation}
\frac{q}{G}
\sim
\left|
1-\frac{a_{\rm G}}{a_{\rm MoS_2}}
\right|
\simeq
\left|
1-\frac{2.46}{3.16}
\right|
\simeq0.22.
\label{eq:tmd_mismatch}
\end{equation}
Therefore
\begin{equation}
\frac{\Lambda_{\rm G/MoS_2}}{\Lambda_{\rm G/hBN}}
\sim
\left(
\frac{0.0168}{0.22}
\right)^2
\simeq
5.8\times10^{-3},
\label{eq:tmd_suppression}
\end{equation}
where the final number retains only the geometric factor, i.e., it sets the ratio of $|V_G|/C_{\rm eff}$ to unity. The factor $5.8\times10^{-3}$ is therefore not a calculated value of $\Lambda_{\rm G/MoS_2}/\Lambda_{\rm G/hBN}$; it states how strongly mismatch alone would suppress reconstruction if the adhesion-to-stiffness ratios were equal. Graphene/TMD and hBN/TMD interfaces are consequently useful large-mismatch controls, consistent with the MoS\(_2\)/graphite and MoS\(_2\)/hBN measurements \cite{Liao2022}. Quantitative prediction still requires their GSFE spectra, elastic tensors, and finite-size depinning flow. Pressure-enhanced adhesion, chemical bonding, edges, or a connected defect network can reverse the low-friction trend, and wear under gigapascal loads is a separate failure mode \cite{Sun2024}.

\section{Discussion}

The framework assigns different jobs to three observables. The directional depinning stress defines the static phase; $\taueq$ and $\Phi_R$ characterize the equilibrium sliding landscape; $\Lambda_n$ measures susceptibility to elastic reconstruction. Translation covariance then fixes the scope of continuum theory: a smooth, clean, infinite moir\'e functional can reconstruct, but it cannot generate an extensive barrier until atomic sampling, commensuration, boundaries, defects, or another covariance-breaking scale is restored.

This statement complements the microscopic finite-contact theories rather than superseding them. The Leven registry index and the Mandelli simulations predict orientation-, load-, and edge-dependent friction in realistic finite flakes \cite{Leven2013,Leven2016,Mandelli2017}. The Yan--Gao analyses resolve shape-dependent static-force oscillations and corner or edge dissipation \cite{YanGao2024Shape,Gao2025Edge}. Equation~\eqref{eq:general_finite_size} places those effects in the same thermodynamic accounting: subextensive boundary forces may dominate an experiment yet vanish after division by area, whereas a nonzero bulk intercept signals a pinned phase. A kinetic-friction exponent, finite-flake crossover, or small corrugation cannot substitute for that intercept.

The graphene/hBN calculation tests the distinction rather than calibrating a universal threshold. Both adhesion models give $\Lambda_{\max}^{\rm test}<0.22$, small in-plane strains, and no resolved physical-coupling barrier in the sampled sector. Replacing the DMC first star by the Fourier-resolved fixed-height Leven landscape increases both $\Lambda$ and the relaxed strain without producing a resolved barrier. Thus the null result is not an artifact unique to the DMC first-star truncation within the tested fixed-height, in-plane models. It does not rule out a change caused by out-of-plane relaxation, nonlinear elasticity, defects, or a denser angular limit. For any fixed loading direction $\hat{\mathbf e}$, a converged nonzero $\liminf_{A\to\infty}\taudep(A,\hat{\mathbf e})$ in such a calculation would overturn the present sliding-side classification without altering the criterion.

Translation covariance supplies an exact counterexample to any universal phase rule based on a bare value of $\Lambda$: within the smooth incommensurate continuum, the adhesion can be scaled so that $\Lambda$ is arbitrarily large while the translated family of relaxed states retains zero bulk barrier. The discrete strong-coupling scan gives a separate numerical illustration. The free/free model crosses $\Lambda=1$ near $g=7.14$, but no size-independent barrier appears before local strains leave the harmonic regime; barriers seen in the smallest rational approximant fall below resolution at larger $N$. The latter observation is not a critical-point determination. Neither a single commensurate supercell nor the equation $\Lambda=1$ fixes $g_c$. Locating a controlled microscopic critical point requires converged adhesion harmonics, out-of-plane relaxation, nonlinear intralayer elasticity, and finite-size continuation of the tilted enthalpy.

The material statement is deliberately narrower than an experimental claim. It applies at zero temperature to defect-free, periodic, in-plane graphene/hBN models over the tested sizes and directions. Finite contacts can still show measurable static or kinetic friction from edges, contamination, grain boundaries, external defects, and load-induced damage \cite{Wang2020Characterization,Qu2020,Hu2024,Sun2024}. Those mechanisms enter the same phase test, but they define different finite-size and disorder ensembles.

\section{Conclusions}

We have defined the static phases of a two-dimensional crystalline interface by the thermodynamic limits of its directional depinning stress. The tensor quantity $\Lambda_n=|V_n|G_{n,i}[D_{\rm rel}^{-1}(\mathbf q_n)]_{ij}G_{n,j}$ instead measures reconstruction; in the nearly aligned hexagonal limit it retains both longitudinal and transverse elastic channels. Translation covariance explains why these are different questions: a clean, smooth, incommensurate continuum can reconstruct without acquiring an extensive sliding barrier.

For ideal graphene/hBN, the DMC first-star model gives $\Lambda_0=0.1401$ for two free layers and $0.06360$ with rigid hBN. The Fourier-resolved fixed-height Leven potential raises the largest tested values to $0.2118$ and $0.09616$. Neither model develops a resolved physical-coupling barrier across three mismatch approximants and five directions; two-dimensional shift-torus scans and multi-start searches for both landscapes likewise find no resolved corrugation or metastable branch. Enhanced coupling crosses $\Lambda=1$ without yielding a controlled bulk threshold before harmonic elasticity fails. Within their stated zero-temperature, periodic, in-plane scope, the tested finite-size sequences are therefore consistent with an elastically relaxed sliding regime. They do not establish zero friction for finite, defective, loaded, thermally activated, or fully three-dimensional interfaces.

Materials screening should follow the same two-step logic: use geometry, elasticity, and the GSFE spectrum to rank reconstruction susceptibility, then assign the phase from finite-size constrained corrugation and quasistatic depinning. A reconstruction threshold, including the one-dimensional Pokrovsky--Talapov value or $\Lambda=1$, is not a substitute for the second step.

\section*{Data and code availability}

All source code and machine-readable data needed to reproduce the reported graphene/hBN values, the DMC--Leven comparison, and the finite-size stress analysis accompany this submission in the \texttt{calculations} and \texttt{calculation\_results} directories. The execution order, software versions, and output-to-figure map are documented in \texttt{README\_REPRODUCIBILITY.md}. These files will also accompany the corresponding arXiv version. No proprietary data were used.

\bibliographystyle{apsrev4-2}
\bibliography{superlubricity-v17}

\end{document}